\documentclass[preprint2]{aastex}
\usepackage[latin1]{inputenc}
\usepackage{graphicx}
\graphicspath{ {.},{./graphics/} }
\usepackage{amsmath}

\DeclareMathOperator*{\argmin}{arg\,min}

\slugcomment{}

\shorttitle{}
\shortauthors{Guennou et al.}

\begin{document}


\title{On the Accuracy of the Differential Emission Measure Diagnostics of Solar Plasmas. Application to AIA / {\it \textbf{SDO}}. Part I: Isothermal plasmas.\\}


\author{C. Guennou\altaffilmark{1}, F. Auch\`ere\altaffilmark{1}, E. Soubri\'e\altaffilmark{1} and K. Bocchialini\altaffilmark{1}}
\affil{Institut d'Astrophysique Spatiale, B\^atiment 121, CNRS/Universit\'e Paris-Sud, 91405 Orsay, France}
\email{chloe.guennou@ias.u-psud.fr}

\author{S. Parenti\altaffilmark{2}}
\affil{Royal Observatory of Belgium, 3 Avenue Circulaire, B-1180 Bruxelles, Belgium}

\and

\author{N. Barbey\altaffilmark{3}}
\affil{SAp/Irfu/DSM/CEA, Centre d'\'etudes de Saclay, Orme des Merisiers, B\^atiment 709, 91191 Gif sur Yvette, France}

\affil{Accepted for publication in The Astrophysical Journal Supplements 2012 September 5.}


\begin{abstract}
DEM analysis is a major diagnostic tool for stellar atmospheres. But both its derivation and its interpretation are notably difficult because of random and systematic errors, and the inverse nature of the problem. We use simulations with simple thermal distributions to investigate the inversion properties of SDO/AIA observations of the solar corona. This allows a systematic exploration of the parameter space and using a statistical approach, the respective probabilities of all the DEMs compatible with the uncertainties can be computed. Following this methodology, several important properties of the DEM inversion, including new limitations, can be derived and presented in a very synthetic fashion. 

In this first paper, we describe the formalism and we focus on isothermal plasmas, as building blocks to understand the more complex DEMs studied in the second paper. The behavior of the inversion of AIA data being thus quantified, and we provide new tools to properly interpret the DEM. We quantify the improvement of the isothermal inversion with 6 AIA bands compared to previous EUV imagers. The maximum temperature resolution of AIA is found to be 0.03 $\log T_e$, and we derive a rigorous test to quantify the compatibility of observations with the isothermal hypothesis. However we demonstrate limitations in the ability of AIA alone to distinguish different physical conditions.
\end{abstract}

\keywords{Sun: corona - Sun: UV radiation}


\section{Motivation}
\label{sec_1}

The Differential Emission Measure (DEM) diagnostic technique offers crucial information about the thermal structuring of the solar and stellar atmospheres, providing a measure of the temperature distribution of the plasma along the line of sight (LOS). However, to derive the DEM from a set of observations is a complex task, due to the inverse nature of the problem, and the understanding of its robustness and accuracy is still relevant today~\citep[e.g.][]{landi2011, testa2012}. Spectrometers are by nature better suited to DEM analysis than broad band imagers. But, because these latter generally offer a higher signal to noise ratio over a larger field of view (FOV), DEM codes have nevertheless been applied to the three coronal bands of the Extreme-ultraviolet Imaging Telescope (EIT) \citep{delaboudiniere1995} or the Transition Region and Coronal Explorer~\citep[TRACE, ][]{handy1999}. However, these instruments were shown not to constrain the DEM enough to reach conclusive results. In recent years, the multiplication of passbands in instruments such as the X-Ray Telescope (XRT) on {\it Hinode}~\citep{golub2007} and the Atmospheric Imaging Assembly (AIA) telescope~\citep{lemen2012} has brought new prospects to reliably estimate the DEM simultaneously over a large FOV. Case studies of the properties of the inversion using these instruments have been published by e.g.,~\cite{martinez2011} or~\cite{reale2009}.

Building on these results, the central objective of the work presented in this series of papers is to provide a systematic characterization of the DEM reconstruction problem to assess both its accuracy and robustness. Using our technique, the capabilities of a given instrument can be evaluated, and new tools facilitating the DEM interpretation are presented. We illustrate our methodology in the specific case of the six coronal bands of AIA, but the same principle can be applied to any set of broad band or spectroscopic measurements.

Initially introduced for element abundance measurements, then further developed by, e.g., \citet{jeferries1972} and \citet{jordan1976}, the DEM formalism has been extensively used in the past decades, on most types of coronal structures, such as  polar coronal holes~\citep{hahn2011}, polar plumes~\citep[e.g.][]{delzanna2003b}, streamers~\citep[e.g.][]{parenti2000}, prominences~\citep[e.g.][]{wiik1993, parenti2007} quiet sun~\citep[e.g.][]{landi1998, parenti2007}, bright points~\citep{brosius2008} or active regions~\citep[e.g.][]{warren2011}. The thermal structuring of the stellar coronae is also investigated using the DEM analysis~\citep[e.g.][]{sanzforcada2003}. In particular, the DEM is one of the tools commonly used to study the thermal stability of the coronal structures just mentioned, and to diagnose the energy source balancing the observed radiative losses. For example, it can help to discriminate between steady or impulsive heating models predicting different loop thermal structures~\citep[see e.g.][]{klimchuk2006, reale2010, susino2010, winebarger2011}. One of the approaches is to establish the cross field thermal structure of resolved loops which is then compared to the DEM simulated for unresolved  multi-stranded and monolitic loops, impulsively or steadily heated.

But reliably inferring the DEM from observations has proved to be a genuine challenge. The fundamental limitations in the DEM inversion have been discussed by, e.g.,~\citet{jeferries1972, craig1976, brown1991, judge1997}, including measurement noises, systematic errors, the width and shape of the contribution functions, and the associated consequences of multiple solutions and limited temperature resolution. Many DEM inversion algorithms have been proposed to cope with these limitations, each with its own strengths and weaknesses~\citep[e.g.][]{withbroe1975, craig1986, judge1997, landi1997, kashyap1998, mcintosh2000, weber2004, goryaev2010, hannah2012}. In parallel to these developments, authors have been attentive early on to estimate the accuracy of the inversions~\citep[e.g. ][]{dere1978}, eventually comparing several algorithms~\citep[e.g. ][]{fludra1986}.

Due to the intrinsic underconstraint of inverse problems and to the inevitable presence of random and systematic measurement errors, multiple physical solutions consistent with the observations exist, even if mathematical uniqueness and stability can be ensured via, e.g., regularization. It is nevertheless possible to quantify the amount of knowledge, or ignorance, on the physical parameter of interest by rigorously defining levels of confidence in the possible solutions or classes of solutions that can explain the observations within the uncertainties. This is a desirable feature for any inversion scheme if it is to be able, for example, to discriminate or even to define, isothermality and multithermality.

In this perspective, we developed a technique to systematically explore the whole space of solutions, in order to determine their respective probabilities and quantify the robustness of the inversion with respect to plasma parameters, random and systematic errors. We used data simulated with simple DEM forms to systematically scan a wide range of plasma conditions, from isothermal to broadly multithermal, and several inversion hypotheses. Comparing the DEM solutions to the input of the simulations, it is possible to quantify the quality of the inversion. Following this strategy, we are able to completely characterize the statistical properties of the inversion for several parametric DEM distributions. We argue that even though the specifics may vary, the main conclusions concerning the existence of multiple solutions and the ability to distinguish isothermality from multithermality also apply to more generic forms of DEM distributions.

In this first paper, we focus on the response of AIA to isothermal plasmas. The properties of the isothermal inversion thus observed will serve as building blocks for the interpretation of the more complex DEM solutions studied in the second paper (hereafter Paper II). Section \ref{sec_2} describes the general methodology and the practical implementation in the case of AIA, including the data simulation, the inversion scheme, the sources of random and systematic errors, and the different DEM distribution models considered. Results for isothermal plasmas are presented and discussed in Section \ref{sec:iso_response}. A summary introducing the treatment of more generic DEM forms is given in conclusion.

\section{Methodology}
\label{sec_2}

\subsection{DEM formalism}
\label{sub_sec_2_1}
Under the assumption that the observed plasma is optically thin, integration along the line of sight (LOS) of collisional emission lines and continua produces in the spectral band $b$ of an instrument an intensity
\begin{equation}
I_b=\frac{1}{4\pi}\int_0^\infty\! R_b(n_e, T_e)\, n_e^2\, ds
\label{eq_1}
\end{equation}
where $R_b(n_e, T_e)$, the response of the instrument to a unit volume of plasma of electron number density $n_e$ and temperature $T_e$, is given by
\begin{equation}
\begin{split}
R_b(n_e, T_e) = & \sum_{X, l} S_b(\lambda_l)\, A_X\, G_{X,l}(n_e, T_e)\\
            & + \int_0^\infty\! S_b(\lambda)\, G_c(n_e, T_e)\, d\lambda.
\end{split}
\label{eq_2}
\end{equation}
The first term of the right member accounts for each spectral line $l$ of each ionic species $X$ of abundance $A_X$, and the second term represents the contribution of the continua. $S_b(\lambda)$ is the spectral sensitivity of the band~$b$ of the instrument. The respective contribution functions $G_{X,l}(n_e, T_e)$ and $G_c(n_e, T_e)$ of the lines and continua contain the physics of the radiation emission processes~\citep[e.g.][]{mason1994} and can be computed using the relevant atomic data. As long as one considers total line intensities, equations~(\ref{eq_1}) and~(\ref{eq_2}) are generic and apply to imaging telescopes as well as to spectrometers.

Summarizing the original reasoning of~\cite{pottasch1963, pottasch1964}, since the function $R_b(n_e, T_e)$ is generally weakly dependent on the density and is peaked with temperature, $I_b$ gives a measure of $\int_p n_e^2 ds$ where the integration is now limited to the portions $p$ of the LOS where the temperature is such that significant emission is produced. If measurements are available at several wavebands, it is possible to plot $\int_p n_e^2 ds$ as a function of the bands' peak temperatures. Generalizing this logic into a differential form, and assuming that the element abundances are constant, equation~(\ref{eq_1}) can be reformulated as
\begin{equation}
I_b=\frac{1}{4\pi}\int_{0}^{+\infty}\! R_b(T_e)\, \xi(T_e)\, d \log T_e
\label{eq_3}
\end{equation}
where $\xi(T_e)=\overline{n_e^2}(T_e)dp/d \log T_e$ is the DEM, that provides a measure of the amount of emitting plasma as a function of temperature\footnote{The logarithmic scale is justified by the shape of the contribution functions (see Figure~\ref{fig:aia_iso_response}). The DEM can also be defined in linear scale as $\xi(T_e)=\overline{n_e^2}(T_e)dp/d T_e$. There is a factor $d\log T_e/dT_e=1/\ln 10\ T_e$ between the two conventions.}. As demonstrated by~\cite{craig1976}, $\overline{n_e^2}(T_e)$ is the mean square electron density over the regions $dp$ of the LOS at temperature $T_e$, weighted by the inverse of the temperature gradients in these regions. The total Emission Measure (EM) is obtained by integrating the DEM over the temperature
\begin{equation}
EM = \int_{0}^{+\infty}\! \xi(T_e)\, d \log T_e = \int_0^\infty \! n_{e}^{2}\, ds.
\label{eq_4}
\end{equation}
Solving the DEM integral equation~(\ref{eq_3}) implies reversing the image acquisition, LOS integration and photon emission processes to derive the distribution of temperature in the solar corona from observed spectral line intensities. We will now investigate the properties of this inversion.

\subsection{Probabilistic interpretation of the DEM solutions}
\label{sub_sec_2_2}
Let us consider a plasma characterized by a DEM $\xi^P(T_e)$. The corresponding intensities observed in $N_b$ spectral bands are noted $I_b^{obs}(\xi^P)$. In order to solve the DEM inverse problem - estimating $\xi^P$ from the observations - one uses a criterion $C(\xi)$ that defines the distance between the data $I_b^{obs}$ and the theoretical intensities $I_b^{th}(\xi)$ computed using equations~(\ref{eq_2}) and~(\ref{eq_3}) for any DEM $\xi(T_e)$. By definition the DEM $\xi^I(T_e)$ solution of the inversion is the one that minimizes this criterion:
\begin{equation}
\xi^I=\argmin_{\xi} C(\xi).
\label{eq_5}
\end{equation}
Since the $I_b^{obs}$ are affected by measurement noises and the $I_b^{th}$ by systematic errors in the calibration and atomic physics, the inversion can yield different solutions $\xi^I$ of probabilities $P(\xi^I | \xi^P)$ for a given DEM $\xi^P$ of the plasma. Bayes' theorem then gives 
\begin{equation}
P(\xi^P | \xi^I) = \frac{P(\xi^I | \xi^P)P(\xi^P)}{P(\xi^I)},
\label{eq_6}
\end{equation}
which is the conditional probability that the plasma has a DEM $\xi^P$ knowing the result $\xi^I$ of the inversion. $P(\xi^I)=\int P(\xi^I | \xi^P)P(\xi^P)\ d \xi^P$ is the total probability of obtaining $\xi^I$ whatever $\xi^P$. In the Bayesian framework, $P(\xi^P)$ is called the {\it prior}. It is uniformly distributed if there is no {\it a priori} information on the DEM $\xi^P$ of the plasma. Conversely, {\it a priori} knowledge or assumptions on the plasma are represented by a varying $P(\xi^P)$. For example, zero probabilities can be assigned to non physical solutions. 

$P(\xi^P | \xi^I)$ contains all the information that can be obtained from a given set of measurements on the real DEM $\xi^P$ of the plasma and as such, it is a desirable quantity to evaluate. Indeed, if the DEM is to be used to discriminate between physical models, as it is for example the case in the coronal heating debate, finding a solution that minimizes the criterion is necessary, but it is not sufficient. It is also crucial to be able to determine if other solutions are consistent with the uncertainties, what are their respective probabilities, and how much they differ from each other.

In principle, and without {\it a priori} on the plasma, $P(\xi^I | \xi^P)$ and thus $P(\xi^P | \xi^I)$ can be estimated for any minimization scheme using Monte-Carlo simulations~\citep{metropolis1949}. For each $\xi^P$, the $N_b$ observed $I_b^{obs}(\xi^P)$ are simulated using equations~(\ref{eq_2}) and~(\ref{eq_3}) and adding photon and instrumental noises. Systematic errors are incorporated to the $I_b^{th}$ and the resulting criterion is minimized. $P(\xi^I | \xi^P)$ is then evaluated from the $N$ solutions $\xi^I$ corresponding to $N$ realizations of the random variables. But since several $\xi^P$ can potentially yield the same $\xi^I$, the derivation of $P(\xi^P | \xi^I)$ from equation~(\ref{eq_6}) requires to know $P(\xi^I)$, the probability to obtain $\xi^I$ whatever $\xi^P$. This is generally not possible, for it requires the exploration of an infinite number of plasma DEMs.

This is why DEM inversion research often focuses on the minimization part of the problem, $P(\xi^P | \xi^I)$ being supposed to be well behaved because of the proper choice of {\it prior} and the multiplication of passbands or spectral lines. However, $P(\xi^P | \xi^I)$ can be computed if the DEM $\xi^P$ of the plasma can be described by a limited number of parameters. In this case, one can scan the whole parameter space and use the Monte-Carlo simulations to estimate $P(\xi^I | \xi^P)$ for all possible $\xi^P$. The possibility that multiple $\xi^P$ yield an identical inversion solution $\xi^I$ being now taken into account, one can determine $P(\xi^I)$ and thus derive $P(\xi^P | \xi^I)$ from equation~(\ref{eq_6}).

This limitation of the complexity of the DEMs that can be considered corresponds to adopting a non-uniform {\it prior} $P(\xi^P)$, while probabilistic treatments were justly developed with the opposite objective of relaxing such non-physical assumptions~\citep[e.g. the MCMC method of ][]{kashyap1998}. But rather than the development of a generic DEM inversion method, our objective is to study the behaviour of $P(\xi^P | \xi^I)$ in controlled experiments. And if the parameterization is properly chosen, the $\xi^P$ can still represent a variety of plasma conditions, from isothermal to broadly multithermal. In addition, we did not make any assumption on the number and properties of the spectral bands, nor on the definition of the criterion nor on the algorithm used to minimize it. The method described to compute $P(\xi^P | \xi^I)$ can therefore be used to characterize any inversion scheme in the range of physical conditions covered by the chosen $\xi^P$ distributions.

\subsection{Inversion method}
\label{sub_sec_2_3}
Devising an efficient way to locate the absolute minimum of the criterion is not trivial. For example, without further assumption, its definition alone does not guarantee that it has a single minimum, so that iterative algorithms may converge to different local minima depending on the initial guess solution. Furthermore, if the value of the minimum itself is a measure of the goodness of fit, it does not provide information on the robustness of the solution. How well the solution is constrained is instead related to the topography of the minimum and its surroundings; the minimum may be deep or shallow and wide or narrow with respect to the different parameters describing the DEM curve.

The number of DEMs resulting in significantly different sets of intensities within the dynamic range of an instrument is potentially extremely large. However, a systematic mapping of the criterion aimed at revealing its minima and their topography is possible if the search is restricted to a subclass of all possible DEM forms. Indeed, if the DEM is fully determined by a limited number of parameters, one can regularly sample the parameter space and compute once and for all the corresponding theoretical intensities $I_b^{th}(\xi)$. The criterion, i.e. the distance between the $I_b^{th}$ and the measured $I_b^{obs}$, is thus computable as a function of the DEM parameters for any given set of observations. It is then trivial to find its absolute minimum and the corresponding DEM solution $\xi^I$, or to visualize it as a function of the DEM parameters.

\subsection{Implementation}
\label{sub_sec_2_4}
\begin{figure*}
\epsscale{1.7}
\begin{center}
\plotone{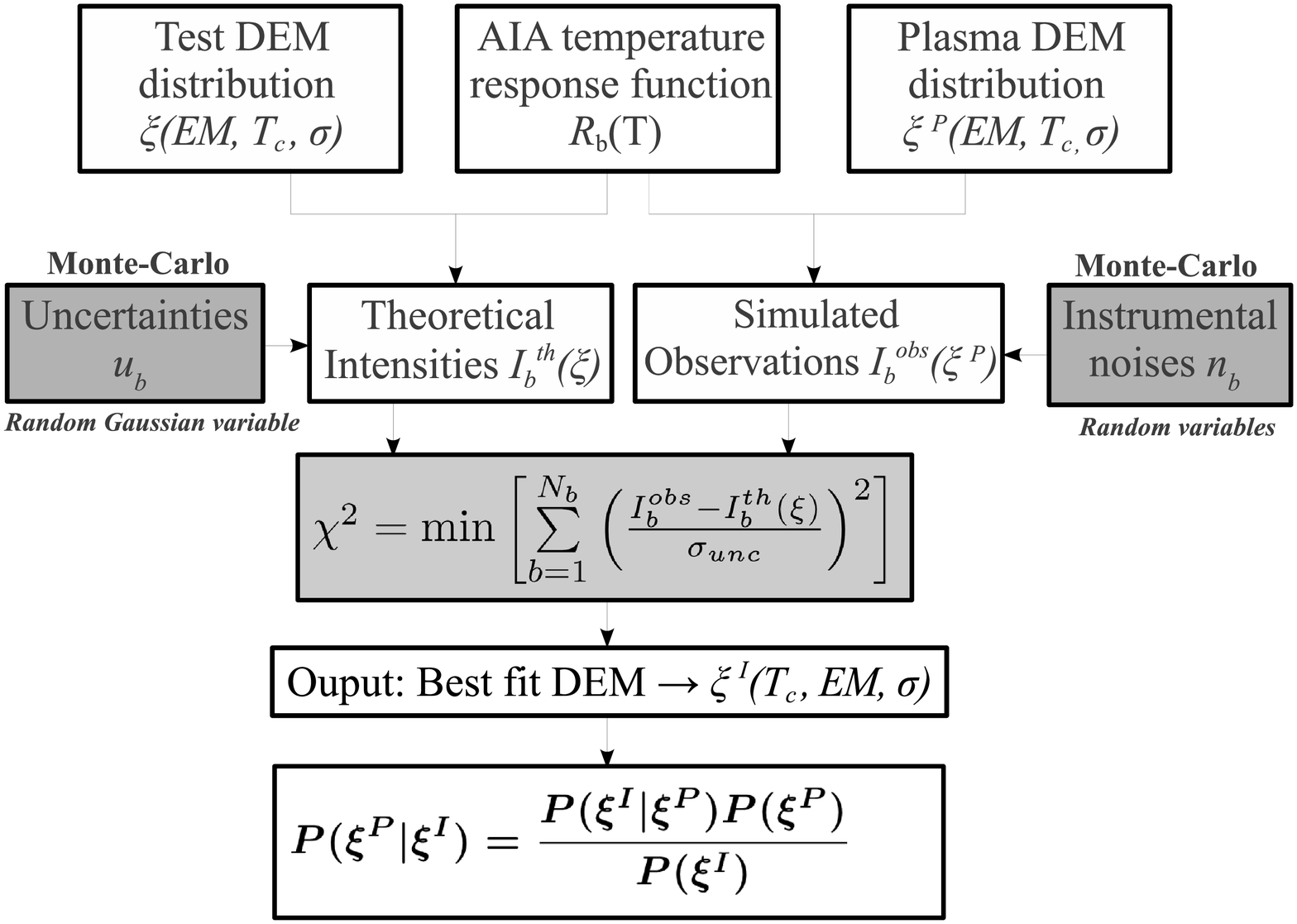}
\end{center}
\caption{Principle of the method used. Reference theoretical intensities $I_b^0$ are tabulated using CHIANTI for different parameterized DEM functional forms (Dirac, Gaussian, top hat). A random variable is added to represent the uncertainties on the calibration and atomic physics. For a given plasma DEM $\xi^P$, AIA observations are simulated in a similar way. A $\chi^2$ criterion is minimized to find the DEM $\xi^I$ that best matches the simulated observations. By scanning the parameters defining $\xi^P$, the probabilities $P(\xi^I|\xi^P)$ and $P(\xi^P|\xi^I)$ are built from a large number of draws of the random variables. These probabilities and the corresponding distributions of $\chi^2$ values give a complete characterization of the inversion for the chosen DEM forms.\label{fig:method}}
\end{figure*}

The procedure used to compute $P(\xi^P | \xi^I)$ is summarized in Figure~\ref{fig:method}.
The parametric DEM forms are described in section~\ref{sec:dem_models}. The intensities $I_b^{obs}$ observed in $N_b$ bands are the sum of average intensities $I_b^0$ and random perturbations $n_b$ due to photon shot noise and measurement errors
\begin{equation}
I_b^{obs} = I_b^0 + n_b.
\label{eq:iobs}
\end{equation}
The $I_b^0$ are equal to the theoretical intensities $I_b^{th}$ in the case of a hypothetically perfect knowledge of the instrument calibration and atomic physics. In practice however, the $I_b^{th}$ are affected by systematic errors $s_b$
\begin{equation}
I_b^{th} = I_b^0 + s_b.
\label{eq:ith}
\end{equation}
Since there is no way of knowing whether the intensities that can be computed from equations~(\ref{eq_2}) and~(\ref{eq_3}) for any DEM $\xi$ are overestimated or underestimated, we identify them\footnote{It is also possible to adopt the view that the intensities computed with CHIANTI are one of the possible estimates of the $I_b^{th}$, in which case we obtain the $I_b^0$ by adding systematic errors. The only difference between the two conventions is the sign of $s_b$. The criterion and therefore the results are identical in both cases.} to the reference theoretical intensities $I_b^0$. The distributions of random and systematic errors are discussed in section~\ref{sec:uncertainties}. The detail of the calculation of the $I_b^0$ is given in section~\ref{sec:theoretical_intensities}. From these, we can either simulate observations $I_b^{obs}$ by adding measurement noises $n_b$ (equation~(\ref{eq:iobs})), or obtain various estimates of the $I_b^{th}$ by adding perturbations representing the systematics $s_b$ (equation~(\ref{eq:ith})).

The criterion $C(\xi)$ and the corresponding minimization scheme are described in section~\ref{sec:criterion}. For any plasma DEM $\xi^P$, Monte-Carlo realizations of the noises $n_b$ and systematics $s_b$ yield several estimates $\xi^I$, from which we compute $P(\xi^I | \xi^P)$. Finally, $P(\xi^P | \xi^I)$ is obtained after scanning all possible plasma DEMs (section~\ref{sec:monte-carlo}).

\subsubsection{DEM distribution models}
\label{sec:dem_models}
Ensuing the discussions of sections~\ref{sub_sec_2_2} and~\ref{sub_sec_2_3}, the $\xi^P$ and $\xi^I$ are both constrained to belong to one of the three following classes of DEM distributions defined by two or three parameters:
\begin{itemize}
\item Isothermal
\begin{equation}
\xi_{iso}(T_{e}) = EM\ \delta (T_e - T_c),
\label{eq:dem_iso}
\end{equation}
where the DEM is reduced to a Dirac $\delta$ function centred on the temperature $T_c$. $EM$ is the total emission measure defined by equation~(\ref{eq_4}).
\item Gaussian in $\log T_e$
\begin{equation}
\begin{split}
\xi_{gau}(T_e) = & EM\mathcal{N}(\log T_e - \log T_c),\\
\mbox{with } \mathcal{N}(x) = & \frac{1}{\sigma\sqrt{2\pi}} \exp\left(-\frac{x^2}{2\sigma^2}\right)
\end{split}
\label{eq:dem_gauss}
\end{equation}
The plasma is here predominantly distributed around a central temperature $T_c$ with a width $\sigma$.
\item Top hat in $\log T_e$
\begin{equation}
\begin{split}
\xi_{hat}(T_e) = & EM\ \Pi(\log T_e - \log T_c),\\
\mbox{with } \Pi(x) = & \begin{cases}
                             \frac{1}{\sigma} & \text{if } |x| < \frac{\sigma}{2}\\
                             0       & \text{else}
                             \end{cases}
\end{split}
\label{eq:dem_hat}
\end{equation}
The plasma is uniformly distributed over a width $\sigma$ around $T_c$.
\end{itemize}
There is no reason for the solar plasma to follow one of these distributions, nor are they the only possible choices. But even though they are simple enough to allow a detailed analysis of the properties of the DEM inversion, they can nonetheless represent a variety of plasma conditions. The conclusions drawn can therefore help understand the behaviour of more generic DEM forms. Furthermore, since the class of solution DEMs $\xi^I$ does not have to be the same as that of the plasma DEMs $\xi^P$, it is possible to investigate the impact of a wrong assumption on the shape of the DEM. For example, one can compute $P(\xi^P | \xi^I)$ for isothermal solutions $\xi^I$ while the plasma DEM $\xi^P$ is multithermal (see paper II).

\subsubsection{Reference theoretical intensities}
\label{sec:theoretical_intensities}
Equations~(\ref{eq_2}) and~(\ref{eq_3}) are used to compute the reference theoretical intensities $I_b^0(\xi)$ for any DEM $\xi$. They are then used to form both simulated observations and various estimates of the theoretical intensities with equations~(\ref{eq:iobs}) and~(\ref{eq:ith}).

From equations~(\ref{eq:dem_iso}),~(\ref{eq:dem_gauss}) and~(\ref{eq:dem_hat}), we derive the expressions of these reference intensities as a function of the parameters $EM$, $T_c$ and $\sigma$ for the three types of DEM distributions.
\begin{itemize}
\item{Isothermal}
\begin{equation}
\begin{split}
I_b^0(EM, T_c) & = EM\int_{0}^{+\infty}\! R_b(T_e)\, \delta(T_e - T_c)\, d\log T_e\\
         & = EM\,R_b(T_c)
\end{split}
\label{eq:ith_iso}
\end{equation}
\item{Gaussian}
\begin{equation}
\begin{split}
I_b^0(EM, T_c, \sigma) & = EM\int_{0}^{+\infty}\! R_b(T_e)\, \mathcal{N}(\log T_e - \log T_c)\, d\log T_e\\
                  & = EM\, (R_b \ast \mathcal{N})(T_c, \sigma)
\end{split}
\label{eq:ith_gauss}
\end{equation}
\item{Top hat}
\begin{equation}
\begin{split}
I_b^0(EM, T_c, \sigma) & = EM\int_{0}^{+\infty}\! R_b(T_e)\,\Pi(\log T_e - \log T_c)\,d\log T_e\\
                  & = EM\, (R_b \ast \Pi)(T_c, \sigma)
\end{split}
\label{eq:ith_hat}
\end{equation}
\end{itemize}
We note that in all cases, the reference theoretical intensities are equal to the convolution product of the instrument response function $R_b(T_e)$ by the chosen DEM $\xi(T_e)$. The $I_b^0$ are pre-computed for all possible combinations of parameters $EM$, $T_c$, and $\sigma$. The appropriate range and resolution to be used for each parameter can be determined from plausible plasma properties and taking into account the instrument characteristics. 

The responses $R_b(T_e)$ of the six AIA coronal bands are computed using equation~(\ref{eq_2}). The contribution functions $G(T_e)$ are obtained using the version 7.0 of the CHIANTI atomic database~\citep{dere1997, dere2009}. We used the CHIANTI ionization balance and the extended coronal abundances. The summation is extended over the 5 nm to 50 nm spectral range for all bands. The instrument sensitivity $S_b(\lambda)$ is obtained as a function of wavelength in units of $\rm{DN}.\rm{cm}^2.\rm{ph}^{-1}.\rm{sr}^{-1}$ by calling the function \texttt{aia\_get\_response} provided in the AIA branch of the Interactive Data Language (IDL) {\it Solar Software} (SSW) package with the \texttt{/DN}, \texttt{/area} and \texttt{/full} keywords. This function implements the AIA pre-flight calibration as described in~\cite{boerner2012}. Since photon shot noise must be taken into account in the error budget (section~\ref{sec:uncertainties}), the $I_b^0(\xi)$ must be computed for given exposure times and not per second. We used the standard AIA exposures of 2 s for the 17.1 nm and 19.3 nm bands, and 2.9 s for the others. 

The contribution functions are computed using CHIANTI from $\log(T_e)=5$ to $\log(T_e)=7.5$ in steps of $0.005\ \log(T_e)$, oversampling the CHIANTI grid by a factor 10 using cubic spline interpolations. The emission measure varies over a wide range from $10^{25}~\mathrm{cm}^{-5}$ to $10^{33}~\mathrm{cm}^{-5}$ in steps of $0.04 \log(EM)$. The DEM width varies linearly in 80 steps from $\sigma=0$ to $\sigma=0.8~\log(T_e)$. This choice of sampling leads to pre-computing $10^7$ groups of 6 AIA intensities, which represents easily manageable data cubes.

\subsubsection{Uncertainties}
\label{sec:uncertainties}
Uncertainties due to random and systematic errors are at the heart of the problem of the DEM inversion. The two affect the observations and their interpretation in different manners~\citep[see e.g.][]{taylor1997}. Observations are mostly affected by random errors caused by both Poisson photon shot noise and nearly Gaussian detection noises like thermal and read noise. These noises vary randomly from pixel to pixel and from exposure to exposure. On the other hand, the errors made on the calibration and atomic physics systematically skew the interpretation of all observed intensities by the same amount and in the same direction.

It is possible to realistically simulate in the $I_b^{obs}$ the statistical properties of the noises affecting the data. The reference intensities $I_b^0$ have units of Digital Numbers (DN). The number of electrons collected in each pixel over the exposure time is obtained by multiplying these values by the gains (in $\rm{e}^-/\rm{DN}$) of the detectors' analog to digital converters listed in SSW. The number of detected photons is then obtained by dividing the result by the quantum yield of the detector, {\it i.e.} the number of photoelectrons produced per interacting photon\footnote{An approximation of the quantum yield of silicon is given by $h\, c / (3.65\, q\, \lambda)$ where 3.65 is the energy in eV required to create an electron hole pair, $q$ is the elementary charge, $c$ is the speed of light in vacuum and $h$ is Planck's constant. Note that in this calculation we assume that all interacting photons have the same wavelength. However, since the full width at half maximum of the AIA bands is comprised between 0.2 and 1.0 nm, the error made is only a few $10^{-3}$.}. These photon intensities are then perturbed by Poisson noise and converted back to photoelectrons. 22 $\rm{e}^-$ RMS of Gaussian CCD read noise~\citep{boerner2012} are finally added before conversion to DN. 

Determining the statistical properties of the systematic errors is more challenging. The tabulated calibration and atomic physics provides a single estimate of the instrument response $R_b$, but systematics nonetheless have a probability distribution. Indeed, the calibration is the result of laboratory measurements themselves affected by random and systematic errors. If we could recalibrate the instrument a number of times in different facilities we would obtain a distribution of instrumental sensitivities $S_b(\lambda)$, the adopted calibration corresponding to one of them. Likewise, different atomic physics codes will give different estimates of the contribution functions $G(n_e, T_e)$, the CHIANTI output being one of them. It is however difficult to characterize these two probability distributions. They are generally implicitly assumed to be Gaussian and the adopted values to be the most probable. But the distributions may in fact be uniform, or asymmetric, or biased, {\it etc}.

The calibration involves a complex chain of measurements, the uncertainties of which are difficult to track and estimate. After independent radiometric calibrations, comparable EUV instruments on SOHO were found to agree only within about 25\%~\citep{huber2002}. Subsequent comparisons could not resolve the discrepancies nor identify their origin in random errors or biases in the individual calibrations. We can only say that the adopted calibration of every SOHO instrument introduces a systematic error in the data analysis but without being able to tell how much and in what direction. It is likely that inter-calibration between AIA and other instruments would run into similar limitations.

\begin{figure*}
\begin{center}
\epsscale{1}
\includegraphics[scale=1]{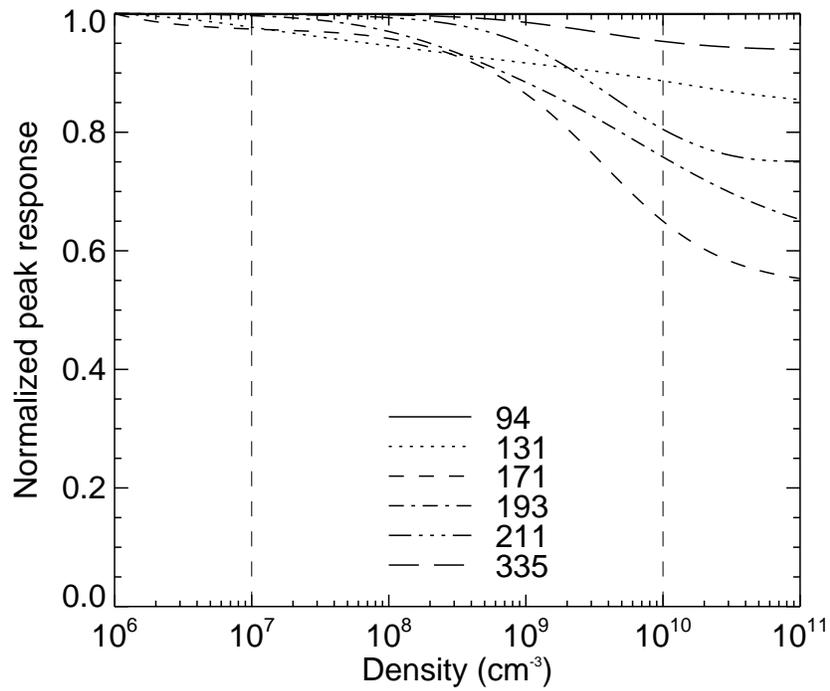}
\end{center}
\caption{Normalized maximum of the response functions $R_b(T_e, n_e)$ of the six AIA coronal bands as a function of electron number density. Only the 9.4 nm band is independent on the density as assumed in the DEM analysis. The other functions vary by up to 35\% in the range of densities plausible in the AIA field of view (dashed vertical lines). This effect induces systematic errors in the DEM inversions.\label{fig:g-vs-n}}
\end{figure*}

Errors in the contribution functions are a major contributor to the uncertainties~\citep[e.g.][]{lang1990, judge1997}. Since the properties of the known atomic transitions are derived either from measurements or modelling, they are not infinitely accurate. Missing transitions lead to underestimated contributions functions, as it is the case for the 9.4 nm channel of AIA~\citep[e.g.][]{aschwanden2011, odwyer2011, foster2011}. The abundances are affected by about 10\% uncertainties~\citep{asplund2009}, not taking into account possible local enhancements of high FIP elements~\citep{young2005}. These imply that, at least in some cases, the abundances are not constant along the line of sight, as assumed in the DEM analysis. The plasma may not be in ionization balance, in which case the CHIANTI calculations of transition rates are not valid. The response functions $R_b$ of AIA are also not independent from the electron number density, which is one of the assumptions made in deriving the DEM expression from equations~(\ref{eq_1}) to~(\ref{eq_3}). When using spectrometers, the spectral lines are chosen so that this hypothesis is effectively verified. We plot in Figure~\ref{fig:g-vs-n} the normalized maximum of $R_b(T_e, n_e)$ versus electron number density. In the AIA field of view, $n_e$ can vary from about $10^{7}\ \rm{cm}^{-3}$ in coronal holes at $1.2\ \rm{R}_\odot$~\citep[e.g.][]{guhathakurta1999} to about $10^{10}\ \rm{cm}^{-3}$ in dense coronal loops~\citep[e.g.][]{reale2002}. In this range, only the 9.4 nm band (solid line) is completely independent on the density. The response function of all other bands decreases as the density increases, the variation reaching about 35\% for the 17.1 nm band (short dashed line). Since the contribution functions have to be computed for a constant electron number density (we chose $10^9\ \mathrm{cm}^{-3}$), they are respectively under or over-estimated if the observed structures are more or less dense. The impact can be mitigated if one has independent knowledge of the range of densities on the LOS, but it nonetheless represents an additional source of uncertainty compared to using density insensitive spectral lines. Finally, these various sources of uncertainties do not affect all spectral bands by the same amount.

Rigorously estimating the properties of the probability distributions of the systematic errors would thus require a detailed analysis of the calibration process and of the atomic physics data and models that is beyond the scope of this paper. In these conditions, we make the simplifying assumption that all systematics are Gaussian distributed and unbiased. According to \citet{boerner2012}, uncertainties on the pre-flight instrument calibration are of the order of 25\%. This is thus interpreted as a Gaussian probability distribution centred on the tabulated values with a 25\% standard deviation. Likewise, we used 25\% uncertainty on the atomic physics for all bands, typical of the estimates found in the literature. Calibration and atomic physics uncertainties were added quadratically for a net 35\% uncertainty on the response functions $R_b$. The $I_b^{th}$ are thus obtained by adding Gaussian random perturbations to the $I_b^0$.

\subsubsection{Criterion and minimization}
\label{sec:criterion}
Since instrumental noises and systematic errors are assumed to be Gaussian distributed, we use a least square criterion
\begin{equation}
C(\xi) = \sum_{b=1}^{N_b} \left( \frac{I_b^{obs} - I_b^{th}(\xi)}{\sigma_b^u}\right)^2
\label{eq:criterion}
\end{equation}
normalized to the total standard deviation $\sigma_b^u$ of the uncertainties in each band. $\sigma_b^u$ is obtained by summing quadratically the standard deviations of the four individual contributions: photon noise, read noise, calibration and atomic physics (section~\ref{sec:uncertainties}). The value of the minimum of $C(\xi)$ corresponding to the solution $\xi^I$ is noted
\begin{equation}
\chi^2 = \rm{min}\ C(\xi).
\label{eq:chi2}
\end{equation}

From equations~(\ref{eq:ith}) and~(\ref{eq:iobs}) we get
\begin{equation}
C(\xi) = \sum_{b=1}^{N_b} \left( \frac{I_b^0(\xi^P) - I_b^0(\xi) + n_b - s_b}{\sigma_b^u}\right)^2
\label{eq:criterion_dev}
\end{equation}
\noindent
If the family of solutions (Dirac, Gaussian or top hat) is identical to that of the plasma DEM $\xi^P$, then in the absence of noise $\chi^2=0$ and the solution $\xi^I$ given by equation~(\ref{eq_5}) is strictly equal to $\xi^P$. However, in the presence of random and systematic errors or if the assumed DEM form differs from that of the observed plasma, $\chi^2$ is not likely to be zero and the corresponding $\xi^I$ may be different from $\xi^P$, for random fluctuations of $n_b$ and $s_b$ can compensate a difference between $I_b^0(\xi^P)$ and $ I_b^0(\xi)$. As discussed in section~\ref{sec:fit_test}, properly interpreting the value of $\chi^2$ provides a means of testing the pertinence of a given DEM model.

Folding equation~(\ref{eq:ith_iso}),~(\ref{eq:ith_gauss}) or~(\ref{eq:ith_hat}) into equation~(\ref{eq:criterion}), we obtain the expression of $C(\xi)$ for the corresponding DEM distributions. Given a set of observed intensities and a DEM model, the criterion can therefore be easily computed for all possible combinations of the parameters $EM$, $T_c$, and $\sigma$ using the $I_b^0(\xi)$ tabulated as described in section~\ref{sec:theoretical_intensities}. Finding its minimum and thus the solution $\xi^I$ is simplified to the location of the minimum of the $C(\xi)$ matrix. This minimization scheme is not fast compared to, e.g., iterative gradient algorithms, but it ensures that the absolute minimum of the criterion is found whatever its topography. Furthermore, this operation can be efficiently implemented on the Graphics Processing Units (GPU) of modern graphics cards by using their CUDA capability. We implemented a scheme in which each GPU core is in charge of computing an element of the $C(\xi)$ matrix, with all GPU cores running in parallel. The search of the minimum of $C(\xi)$ is also performed by the GPU, thus reducing the transfers between GPU to CPU to the values of $\chi^2$ and $\xi^I$.

\subsubsection{Monte-Carlo simulations}
\label{sec:monte-carlo}
Restricting $\xi^P$ and $\xi^I$ to belong to one of the DEM classes described in section~\ref{sec:dem_models}, $P(\xi^I | \xi^P)$ and $P(\xi^P | \xi^I)$ are evaluated from Monte-Carlo simulations. For every combination of the two or three parameters defining $\xi^P$ (the ranges and resolutions being given in~\ref{sec:dem_models}), 5000 independent realizations of the random and systematic errors are obtained. For each of the corresponding sets of six simulated AIA intensities, the inversion code returns the values of the parameters defining $\xi^I$ (equation~(\ref{eq_5})) corresponding to the absolute minimum of the criterion (equation~(\ref{eq:criterion})). From the resulting 5000 $\xi^I$ we estimate the conditional probability $P(\xi^I | \xi^P)$ with a resolution defined by the sampling of the parameters. Integration over $\xi^P$ gives $P(\xi^I)$ and using Bayes' theorem we obtain $P(\xi^P | \xi^I)$.

\section{Results: isothermal solution to isothermal plasma}
\label{sec:iso_response}
In order to understand the fundamental properties of the DEM inversion of the AIA data, we first applied the method to investigate the behaviour of the isothermal solutions to simulations of isothermal plasmas. The electron temperatures and emission measures of the plasmas are noted $T_c^P$ and $EM^P$ respectively. The corresponding inverted quantities are noted $T_c^I$ and $EM^I$. The probabilities $P(T_c^I, EM^I | T_c^P, EM^P)$ and $P(T_c^P, EM^P | T_c^I, EM^I)$ are stored in matrices of dimension 4. To maximize the clarity of the results, and since the thermal content of the plasma is the main object of DEM analysis, we reduce the number of dimensions by fixing the emission measure of the simulated plasmas to be $EM^P=2\times 10^{29}\ \mathrm{cm}^{-5}$. Furthermore, the probabilities are always presented whatever the emission measure by integrating them over $EM^I$, even though $EM^I$ is of course solved for in the inversion process. 

\begin{figure*}
\epsscale{1}
\begin{center}
\includegraphics[scale=0.7]{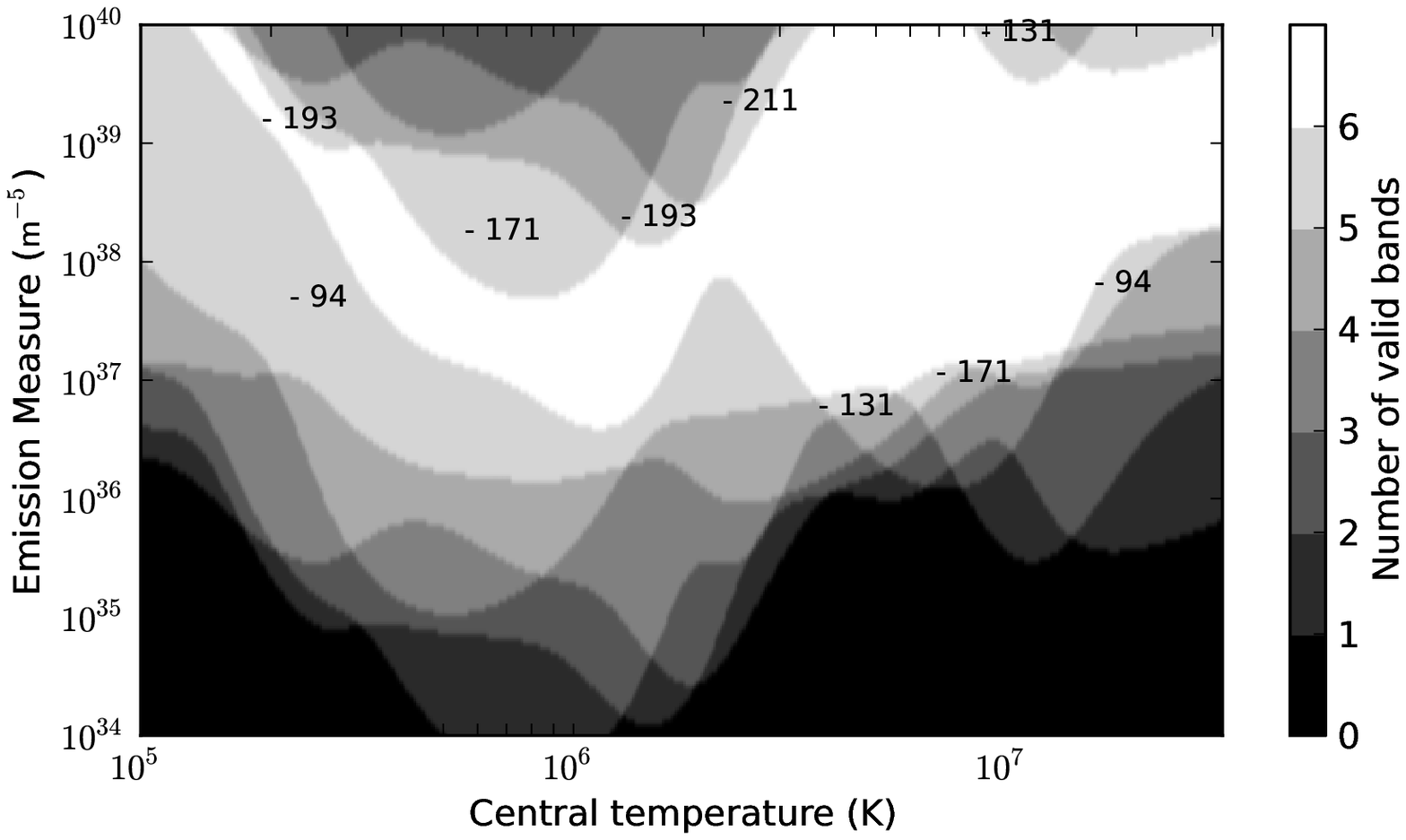}
\includegraphics[scale=0.7]{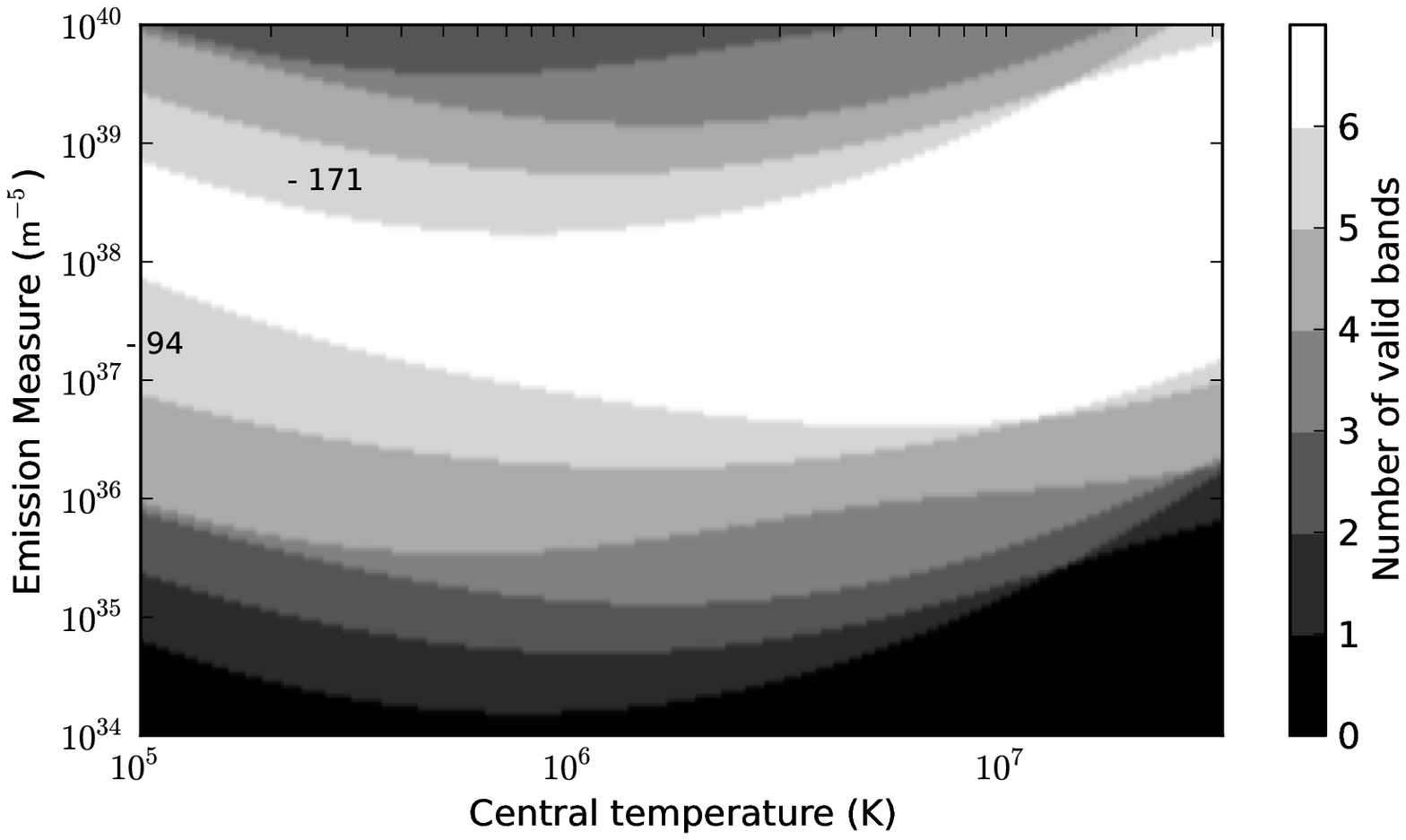}
\end{center}
\caption{Number of AIA bands in which the signal is comprised between 1 DN (detection threshold) and 11000 DN (saturation) as a function of temperature and emission measure, for standard exposure times. Left: isothermal plasmas ; right: Gaussian DEMs with $\sigma = 0.5\ \log (T_e)$. Only solar structures falling in the white regions produce exploitable signal in all six AIA coronal bands. The regions corresponding to five valid bands are labelled with the wavelength of the missing one. If exposure times were increased, the boundaries of all regions would be shifted towards smaller emission measures. If several images were summed up to overcome saturation, the upper boundaries would be moved upwards.\label{fig:valid_bands_iso}}
\end{figure*}

The chosen $EM^P$ is typical of non flaring active regions~\citep[e.g.][]{warren2011}. Figure~\ref{fig:valid_bands_iso} shows as a function of $T_c^P$ and $EM^P$ the number of AIA bands in which a plasma produces more than 1~DN (detection threshold) and less than 11000~DN (saturation). The left panel is for isothermal plasmas, the right panel for Gaussian DEMs with $\sigma^P=0.5 \log(T_e)$. At the chosen $EM^P$, and since we did not implement the detector saturation in our simulations, we always have exploitable signal in all six AIA coronal bands, except below a few $10^5$~K. Conversely, solar structures outside the white areas produce signal only in some of the six bands, unless spatial or temporal summation is used. Therefore, the results presented in the following sections correspond to optimum conditions outside of which the combination of higher noise and possible lower number of valid bands will always lead to weaker constraints on the DEM.

\subsection{Three bands: EIT, TRACE, or low emission measures}
\label{sec:3bands}

We first present inversion results using only three bands as an illustration of the situation encountered with previous EUV imaging telescopes like EIT, TRACE or EUVI. The 17.1~nm and 19.5~nm coronal passbands of EIT and TRACE have direct equivalents in AIA, but the \ion{Fe}{15} 28.4~nm band does not. After comparison of its isothermal response~\citep[see, e.g., Figure 9 of][]{delaboudiniere1995} with those of AIA (Figure~\ref{fig:aia_iso_response}), we chose the \ion{Fe}{14} 21.1~nm band as its closest AIA counterpart. The three bands configuration is also similar to having six bands and a low emission measure plasma\footnote{For completeness, the plots for all combination of three to six bands are available on line at ftp.ias.u-psud.fr/cguennou/DEM\_AIA\_inversion/}. Indeed, at $5\times 10^{26}\ \mathrm{m}^{-5}$ and $1.5\times 10^6$~K, values typical of coronal loops, only three of the six AIA coronal bands produce more than 1~DN (see Figure~\ref{fig:valid_bands_iso}), the others providing only upper limit constraints to the DEM. 

\begin{figure*}
\epsscale{1}
\begin{center}
\includegraphics[scale=1]{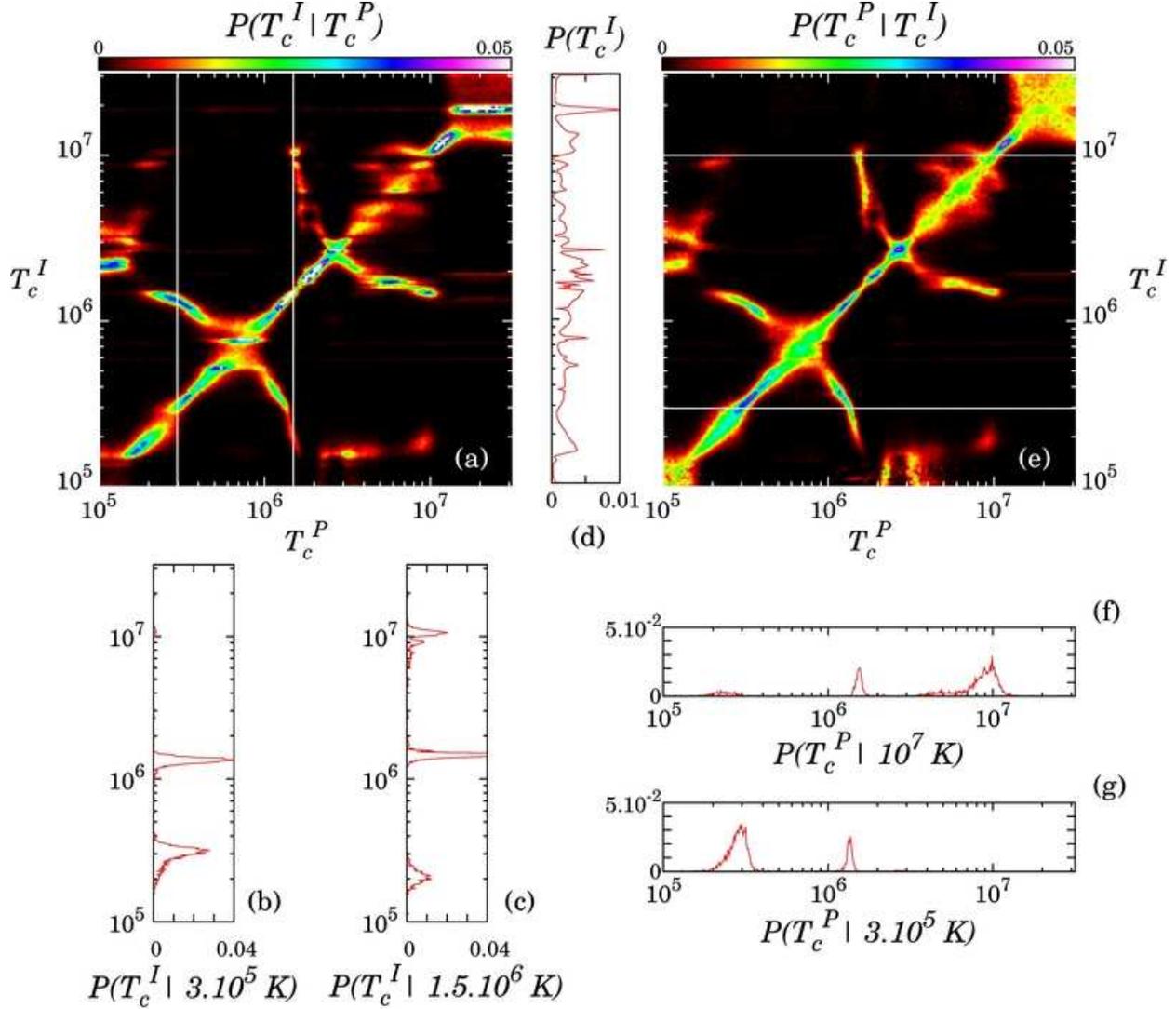}
\end{center}
\caption{Probabilities of the isothermal solutions for observations of an isothermal plasma with three of the AIA coronal bands (17.1, 19.3 and 21.1 nm). (a): reading vertically, conditional probability $P(T_c^I|T_c^P)$ that the inversion yields $T_c^I$ for a given plasma temperature $T_c^P$. (e): reading horizontally, probability  $P(T_c^P|T_c^I)$ that the plasma has a temperature $T_c^P$ for an inverted value $T_c^I$. (e) is obtained by normalizing (a) to (d) the unconditional probability $P(T_c^I)$ that the inversion yields $T_c^I$ whatever the plasma temperature, which is obtained by integrating (a) over $T_c^P$. The branches bifurcating from the diagonal reveal the existence of multiple solutions. The probability profiles (b) and (c) show for example that $3\times 10^5$~K or $1.5\times 10^6$~K plasmas can be measured at $3\times 10^5$~K, $1.5\times 10^6$~K or $10^7$~K. {\it Vice versa}, the profiles (f) and (g) can be used to properly interpret $3\times 10^5$~K and $10^7$~K inversions as also both compatible with a $1.5\times 10^6$~K plasma. See section~\ref{sec:3bands} for details.\label{fig:3bands}}
\end{figure*}

Panel (a) of Figure~\ref{fig:3bands} shows a map of the probability\footnote{Defined as the probability for the solutions to lie between  $\log T_c$ and $\log T_c + \Delta \log T_c$. } $P(T_c^I | T_c^P)$. It is worth noting that, as explained in section~\ref{sub_sec_2_2}, $P(T_c^I)$ and thus $P(T_c^P | T_c^I)$ could be evaluated only because the limitation to simple parameterized DEM forms allowed the computation of $P(T_c^I | T_c^P)$. The plot of $P(T_c^I)$ (and thus the horizontal structures in panel (a)) shows that some temperature solutions $T_c^I$ are more probable than others for any plasma temperature $T_c^P$. In the case of real observations, this can be misinterpreted as the ubiquitous presence of plasma at the most likely temperatures. This caveat was already analysed by~\cite{weber2005} in the case of the 19.5 to 17.3 nm TRACE band ratio and we will discuss it further in Paper II for multithermal plasmas.

\begin{figure*}
\epsscale{1}
\begin{center}
\includegraphics[scale=1]{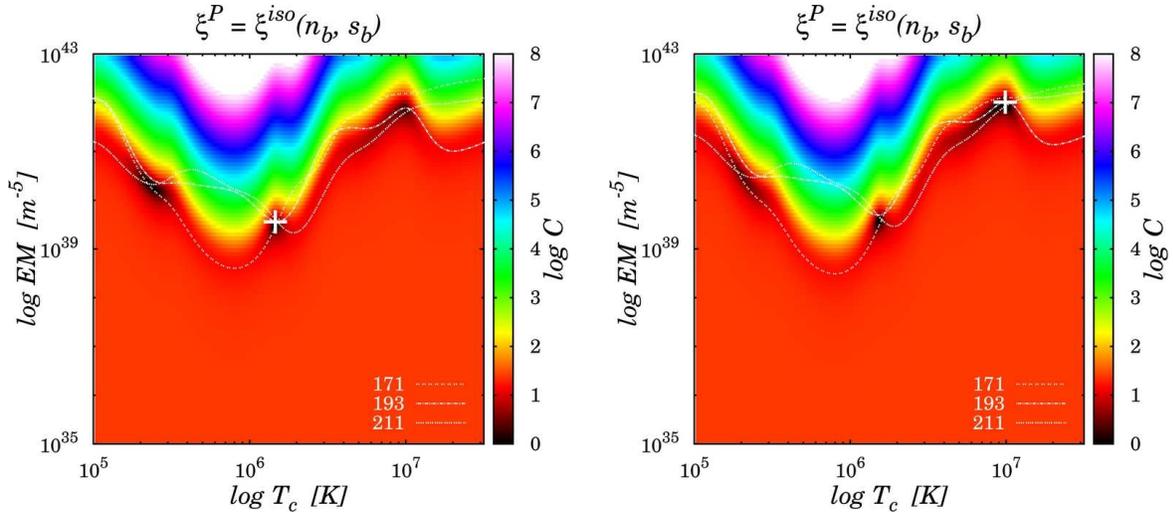}
\end{center}
\caption{The least squares isothermal criterion (Equation~(\ref{eq:criterion})) for a simulated isothermal plasma at $T_c^P=1.5\times 10^6$~K. The two panels correspond to two of the random draws used to build Figure~\ref{fig:3bands}. The loci curves for the three components of the criterion are superimposed, and the white plus signs mark the location of its absolute minimum. Both solutions are fully consistent with the simulated data given the uncertainties.\label{fig:criterion}}
\end{figure*}

Both probability maps exhibit a diagonal from which several branches bifurcate. Below $2\times 10^5$~K and above $10^7$~K the diagonal disappears because since the bands have little sensitivity in these regions, the signal is dominated by noises and the inversion output is thus independent from the temperature. The general symmetry with respect to the diagonal reflects the equality $P(T_c^P | T_c^I)=P(T_c^I | T_c^P)/P(T_c^I)$. The diagonal is formed by the solutions $T_c^I$ that are close to the input $T_c^P$, while the branches correspond to significant deviations from the input. In $P(T_c^I | T_c^P)$, these branches imply that two or more solutions $T_c^I$ can be found for a same plasma temperature $T_c^P$. Conversely, reading horizontally the $P(T_c^P | T_c^I)$ image, a given temperature solution $T_c^I$ can be coherent with two or more plasma temperatures $T_c^P$. The (b) and (c) plots give the probability of the solutions for two plasma temperatures. At $T_c^P=3\times 10^5$~K, the solution may be $T_c^I=3\times 10^5$~K or $1.2\times 10^6$~K. At the typical coronal temperature $T_c^P=1.5\times 10^6$~K, the inversion can yield $1.5\times 10^6$~K but also $2\times 10^5$~K or $10^7$~K. It is thus possible to incorrectly conclude to the presence of cool or hot coronal plasma while observing an average million degree corona. This ambiguity has far reaching implications since the detection of hot plasma is one of the possible signatures of nano-flares~\cite[e.g.][]{cargill1994, klimchuk2006}. Since by definition they correspond to the absolute minimum of the criterion, all solutions are fully consistent with the data given the uncertainties. One or more of the multiple solutions can be rejected only based on additional independent {\it a priori} information. For example, the high temperature solution corresponds to an emission measure of $4\times 10^{31}\ \mathrm{cm}^{-5}$ (right panel of Figure~\ref{fig:criterion}), which is extremely high considering the present knowledge of the corona. If no such information is available however, both low and high temperature solutions can still be correctly interpreted as also compatible with a $1.5\times 10^6$~K plasma with the aid of the $P(T_c^P | 3\times 10^5)$ and $P(T_c^P | 10^7)$ probability profiles (f) and (g).

The reason for the formation of these branches is illustrated by Figure~\ref{fig:criterion}. On both panels, the background image is the value of the criterion $C(\xi)$ for a $T_c^P=1.5\times 10^6$~K plasma as a function of $T_c$ and $EM$. The absolute minimum of the criterion, the arguments of which are the inverted parameters $T_c^I$ and $EM^I$ (equation~(\ref{eq_5})), corresponds to the darkest shade of grey and is marked by a white plus sign. The criterion is the sum of three components, one per waveband (equations~(\ref{eq:criterion}) and~(\ref{eq:criterion_dev})). The three superimposed curves are the loci emission measure curves for each band $b$, i.e. the location of the ($T_c$, $EM$) pairs for which the theoretical intensities $I_b^{th}$ equal the measured ones $I_b^{obs}$. Below the loci curves, the criterion is almost flat because at lower emission measures the $I_b^{th}$ are much smaller than the constant $I_b^{obs}$. Conversely, the criterion is dominated by the $I_{th}$ at high emission measures. The darkest shades of gray and thus the minimum of the criterion are located between these two regions. The two panels correspond to two independent realizations of the random and systematic errors. For each draw, the loci curves are randomly shifted along the $EM$ axis around their average position. In the absence of errors, the three loci curves would cross in a single point at the plasma temperature $T_c^P$, giving a criterion strictly equal to zero. In the left panel, with random and systematic errors included, they do not intersect at a single point but the non-zero absolute minimum of $C$, where they are the closest together, is around $T_c^P$. However, the criterion has two other local minima, around $2\times 10^5$~K and around $10^7$~K, where two or three of the loci curves also bundle up. In the right panel, a different random draw shifts the curves closest together around the high temperature local minimum that thus becomes the new absolute minimum. For this $1.5\times 10^6$~K plasma, the inversion thus yields solutions randomly located around the several local minima with respective probabilities given by the profile of Figure~\ref{fig:3bands}(c). When scanning the plasma temperatures, the positions of the minima vary, thus building the branches in the probability maps. In addition, depending on their location the minima can be more or less extended along one or the other axes, which results in a varying dispersion around the most probable solutions.

Systematic errors are simulated with random variables while they are in fact identical for all measurements. Thus, the computed $P(T_c^I|T_c^P)$ does not give the probability of solutions $T_c^I$ for the practical estimates of the calibration and atomic physics. In reality the output of the inversion is biased towards one or the other of the multiple solutions, but we do not know whether the calibration and atomic physics are under or over-estimated. Therefore, in order to deduce the probability that the plasma has a temperature $T_c^P$ from an inverted temperature $T_c^I$, we must account for the probabilities of the systematics as defined in section~\ref{sec:uncertainties}. The randomization samples their distribution, which ensures that the estimated $P(T_c^P|T_c^I)$ are the probabilities relevant to interpret $T_c^I$.

\subsection{Six bands: AIA}
\label{sec:6bands}

\begin{figure*}
\epsscale{1}
\begin{center}
\includegraphics[scale=1]{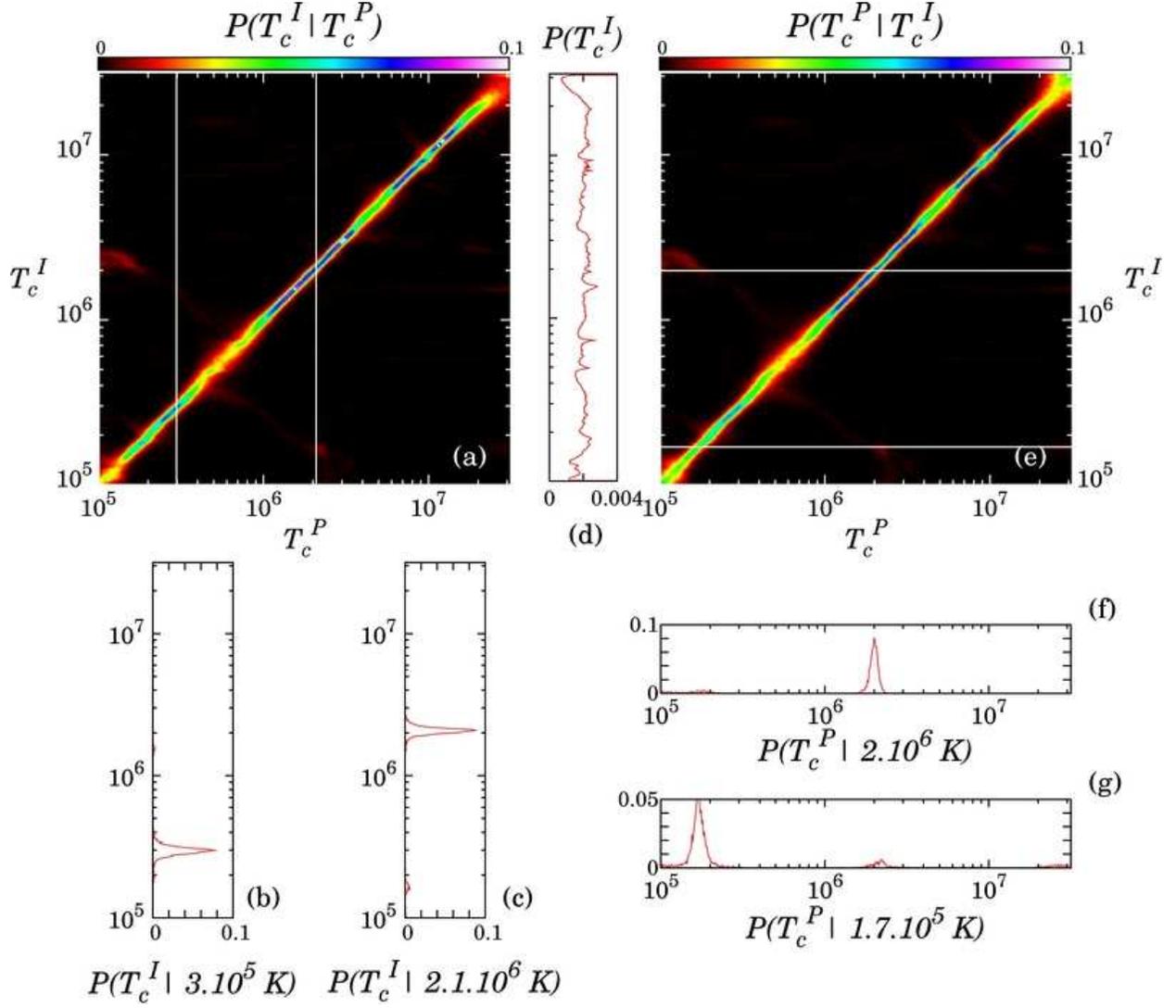}
\end{center}
\caption{Same as Figure~\ref{fig:3bands} for the six AIA coronal bands. The determination of the temperature of the simulated isothermal plasma is now unambiguous. From the width of the diagonal we deduce the resolution of the isothermal inversion to be about $0.05\  \log(T_c)$.\label{fig:6bands}}
\end{figure*}

Figure~\ref{fig:6bands} is the same as Figure~\ref{fig:3bands}, but now including the six AIA coronal bands in the analysis. Some secondary solutions persist at low probabilities but compared to the three bands case, most of the solutions are now concentrated on the diagonal. This illustrates that the robustness of the inversion process increases with the number of bands or spectral lines. Comparison with Figure~\ref{fig:3bands} quantifies the improvement brought by AIA over previous instruments. Neglecting the low probability solutions, if independent {\it a priori} knowledge justifies the isothermal hypothesis, the six AIA bands thus provide an unambiguous determination of the plasma temperature. The temperature resolution of the inversion can be estimated from the width at half maximum of the diagonal. It varies over the temperature range between 0.03 and 0.11 $\log T_c^P$. It is of course be modified if we assumed different uncertainties on the calibration and atomic physics than the ones chosen in section~\ref{sec:uncertainties}. We tested the sensitivity of the temperature resolution to the level of uncertainties $\sigma_b^u$, from 10 to 55\%. The higher the uncertainties, the lower the temperature resolution and the more probable the secondary solutions\footnote{The corresponding probability maps are available on line at ftp.ias.u-psud.fr/cguennou/DEM\_AIA\_inversion/}. For an estimated temperature $T_c^I$ of 1~MK, the temperature resolution of the inversion varies between 0.02 for 10\% error and 0.08 $\log T_c^P$ for 55\% error. In the worst case, for 55\% errors, the temperature resolution decreases to $0.2 \log T_c^P$ for the temperature interval between 0.5 and 0.9 MK. At 1~MK, the resolution is proportional to the uncertainty level with a coefficient of 0.15 ($\Delta T_c^P \sim 0.15 \ \sigma_b^u$).

Since by definition our method always finds the absolute minimum of the least square criterion of equation~(\ref{eq:criterion}), the derived temperature resolution is an intrinsic property of the data and not of the inversion scheme. It is the result of the combination of the random and systematic errors and the shapes the contribution functions. Its value is directly comparable with the findings of~\cite{landi2011}. These authors showed that the temperature resolution of the MCMC code of~\cite{kashyap1998} applied to isothermal plasmas is 0.05 log $T$. Their tests were made on simulated observations of a $10^6$~K plasma in 45 isolated spectral lines with 20\% random errors. Assuming that the MCMC method does converge towards solutions consistent with the limitations of the data, the fact that the temperature resolution is comparable for 6 AIA bands and 45 spectral lines suggests that, in the isothermal case, it is driven by the uncertainties level rather than the number of observables. This conclusion is consistent with the isothermal limit of Figure 6 of ~\cite{landi2010}.

\subsection{Residuals and goodness of fit test\label{sec:fit_test}}

The probability maps presented in the above sections are valid for a given hypothesis on the plasma DEM distribution, but they would be useless without a test of its validity. The pertinence of the DEM model chosen to interpret the observations can be assessed by analyzing  the distribution of the sum of squared residuals defined by equations~(\ref{eq:criterion}) and~(\ref{eq:chi2}). If applying our inversion scheme to real data, we could compare the resulting residuals to the distribution derived from simulations for a given DEM model and thus quantify the probability that the data is consistent with the working hypothesis (e.g. isothermal or Gaussian).

\begin{figure}
\epsscale{1.0}
\begin{center}
\includegraphics[scale=0.7]{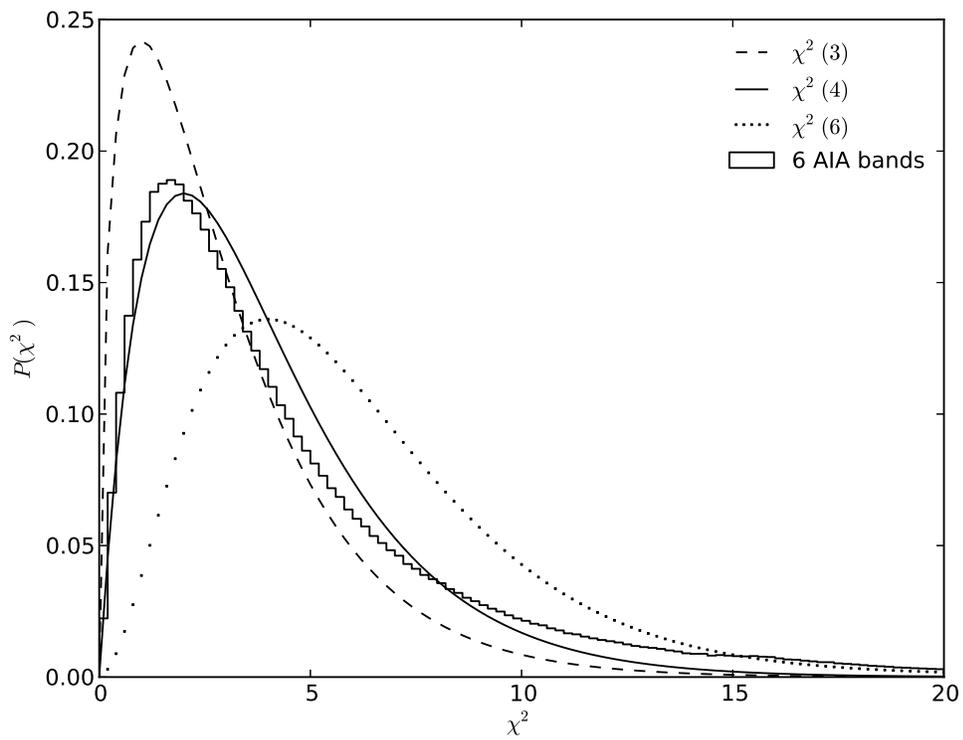}
\end{center}
\caption{The observed distribution of the sum of the squared residuals (solid histogram) \textbf{differs somewhat from the expected degree 4 $\chi^2$} distribution (solid curve). It is slightly shifted toward a degree 3 (dashed curve), which can be explained by a small correlation between the six AIA coronal bands.}\label{fig:chi2_iso}
\end{figure}

\begin{figure*}[ht]
\begin{center}
\includegraphics[scale=1]{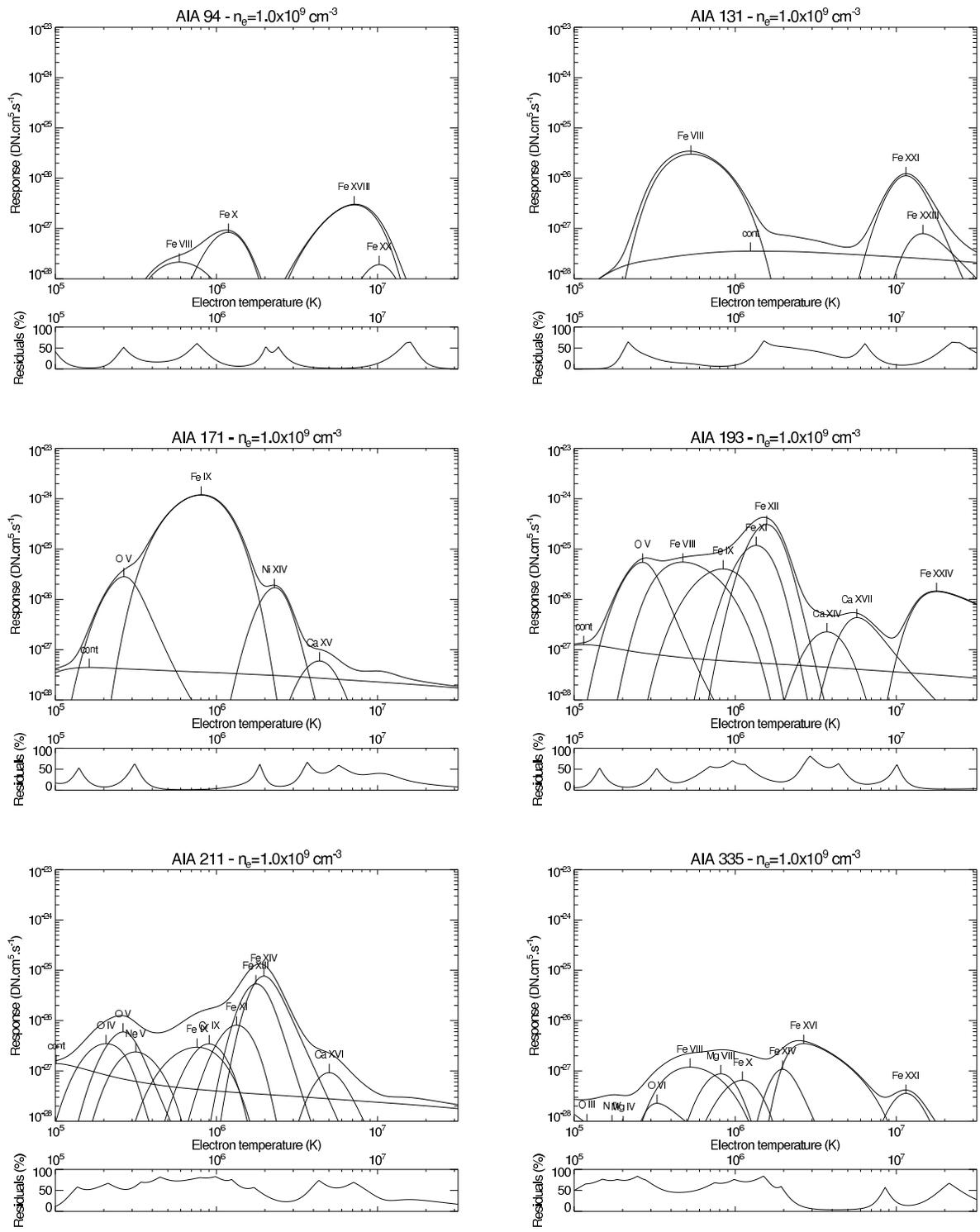}
\end{center}
\caption{Isothermal response of the six AIA coronal bands between $10^5$ and $3\times 10^7$ K. For each band, the thick curve is the total response and the thin curves are the partial responses for the ions that contribute the most at at least one temperature. The fraction of the total response accounted for by those dominant ions is shown below each main plot. Computations for an electron number density of $10^9\ \mathrm{cm}^{-3}$.\label{fig:aia_iso_response}}
\end{figure*}

The solid line histogram of Figure~\ref{fig:chi2_iso} shows the distribution of $\chi^2$ values corresponding to the plots of Figure~\ref{fig:6bands}. The distribution is close to a degree~4 $\chi^2$ distribution (solid curve) although not a perfect match, with a peak shifted to the left and an enhanced wing. The most probable value of the squared residuals is $\sim 1.73$ and $95$\% of them are comprised between 0 and 15. Whatever the actual plasma DEM, any inversion made with the isothermal hypothesis and yielding a $\chi^2$ value in this range can thus be considered consistent with an isothermal plasma given the uncertainties. This isothermality  test is similar to that recommended by~\cite{landi2010}, identifying our $\chi^2$ to their $F_\mathrm{min}$ and our maximum acceptable $\chi^2$ to their $\Delta F$. This does not imply however that isothermality is the only nor the best possible interpretation of the data, for different DEMs can produce similar $\chi^2$ values. The discrimination between DEM models will be discussed in Paper II.
 
The properties of the empirical distribution of squared residuals can be explained as follows. Since we simulated observations of a purely isothermal plasma, an isothermal model can always represent the data. Without errors, there would always be one unique couple ($T_c^I$, $EM^I$) corresponding to six intensities perfectly matching the six AIA observations, thus giving zero residuals. With errors, if we forced the solution ($T_c^I$, $EM^I$) to be the input ($T_c^P$, $EM^P$), the summed squared residuals resulting from a number of random draws should have by definition the Probability Density Function (PDF) of a degree six $\chi^2$ distribution (dotted curve of Figure~\ref{fig:chi2_iso}), for we have six independent values of $I_b^{obs} - I_b^{th}$ and we normalized the residuals to the standard deviation $\sigma_b^u$ of the uncertainties. But since we solve for two parameters ($T_c^I$, $EM^I$) by performing a least squares minimization at each realization of the errors, the solution is not exactly the input ($T_c^P$, $EM^P$) and we should expect a PDF with two degrees of freedom less (dashed curve). Instead of being a pure degree~4, the observed distribution is slightly shifted toward a degree 3 because of two factors. First, the errors are a combination of Poisson photon noise and Gaussian read noise, while the $\chi^2$ distribution is defined for standard normal random variables. Second, as discussed below, the six residuals are not completely independent.

Figure~\ref{fig:aia_iso_response} shows the response functions $R_b(T_e)$ of the  AIA bands to isothermal plasmas with electron temperatures from $10^5$ to $3\times10^7$~K for a constant electron number density of $10^9\ \mathrm{cm}^{-3}$. For each band, the thick curve is the total response, and the labeled thin curves are the partial responses for the ions that contribute the most for at least one temperature. The fraction of the total response not accounted for by those dominant ions is shown below each main plot. Ionization stages common to several bands are found across the whole range of temperatures. \ion{O}{5} dominates the response at $2.5\times 10^5$~K in the 17.1~nm, 19.3~nm and 21.1~nm bands. Around 1~MK, \ion{Fe}{9} is found in the 17.1~nm, 19.3~nm and 21.1~nm responses, and \ion{Fe}{10} contributes to the 94~nm, 21.1~nm and 33.5~nm bands. At 2 MK, \ion{Fe}{14} is common to the 21.1~nm and 33.5~nm bands. This is consistent with the analysis of the AIA bands by~\cite{odwyer2010}. Because of this redundancy, the response functions tend to have similar shapes in the regions of overlap, resulting in a correlation between the residuals. 

\section{Summary and conclusions}

By restricting the solutions to functional forms described by a limited number of parameters, we obtained a complete statistical characterization of the DEM inversion. Even though they are not expected to accurately describe real coronal properties, these simple DEM distributions can nonetheless model a wide range of plasma conditions. The results presented in this series of papers can thus be fruitfully used to demonstrate many important properties and guide the interpretation of the output of generic DEM inversion codes. We illustrated the method by applying it to the six coronal bands of the AIA telescope. In this first paper, we limited ourselves to isothermal plasmas and isothermal solutions.

The case presented in section~\ref{sec:3bands} demonstrates the existence of multiple solutions if the number of bands is limited either by design of the instrument or by lack of signal. However, since our method provides the respective probabilities of the multiple solutions, it is possible to properly interpret the solutions as compatible with several plasma temperatures. Even if some of these properties have been illustrated in case studies, we provide here a systematic analysis of a wide range of plasma parameters. The computed distribution of squared residuals can be used to test the coherence of real AIA data with the isothermal hypothesis. This type of analysis can also be help to determine the optimum data acquisition parameters for AIA (e.g. spatial binning and exposure time) ensuring that no secondary solution is present. In section~\ref{sec:6bands}, we showed that, with enough signal, the six AIA coronal bands provide a robust reconstruction of isothermal plasmas with a temperature resolution comprised between 0.03 and 0.11 $\log T_e$. The comparison with the three bands case gives a quantification of the improvement brought by the new generation of instruments.  The same method can be applied to other instruments with different response functions and different numbers of bands or spectral lines. This naturally requires the computation of the corresponding probability matrices and distribution of residuals. 

The temperature resolution, and more generally the details of the probability matrices presented in sections~\ref{sec:3bands} and~\ref{sec:6bands}, depend on the amplitude and distribution of the random and systematic errors. We found the resolution to be proportional to the uncertainty level (at 1~MK, $\Delta T_c^P \sim 0.15 \ \sigma_b^u$). We simulated plasmas with high emission measures typical of active regions. Depending on the temperature, either the photon noise or the uncertainties on the calibration and atomic physics dominate. The illustrated properties of the inversion, from the multiplicity of solutions to the temperature resolution, are thus driven by both random systematic errors. While the photon noise can be reduced by increasing the exposure time or binning the data, reducing the systematics requires better atomic data and photometric calibration, which is not trivial. 

\acknowledgments
S.P. acknowledges the support from the Belgian Federal Science Policy Office through the international cooperation programmes and the ESA-PRODEX programme and the support of the Institut d'Astrophysique Spatiale (IAS). F.A. acknowledges the support of the Royal Observatory of Belgium. The authors would like to thank J. Klimchuk for fruitful discussions and comments.

\clearpage

\clearpage

\clearpage

\clearpage

\clearpage

\clearpage

\clearpage

\end{document}